\documentstyle[preprint,eqsecnum,aps]{revtex}

\newtheorem{proposition}{Proposition}
\newtheorem{theorem}[proposition]{Theorem}
\newtheorem{corollary}[proposition]{Corollary}

\newcommand{\bq}{\bbox{q}}
\newcommand{\bp}{\bbox{p}}
\newcommand{\rd}{\partial}

\begin{document}
\draft
\preprint{hep-th/9511027}

\title{Quantum mechanics of higher derivative systems \\
and total derivative terms}
\author{Yasuhito Kaminaga\thanks{Electronic address: kaminaga@toho-u.ac.jp}}
\address{Department of Physics, Toho University \\
Miyama 2-2-1, Funabashi 274, Japan}
\date{\today}
\maketitle

\begin{abstract}
A general theory is presented
of quantum mechanics of singular, non-autonomous, higher derivative systems.
Within that general theory, $n$-th order and $m$-th order Lagrangians
are shown to be quantum mechanically equivalent
if their difference is a total derivative.
\end{abstract}


\section{Introduction}

Higher derivative theories occur in various aspects of
modern physics---gravity, strings, particle phenomenology, and so on.
It is of importance to clarify general properties of such theories.

The purpose of this paper is to prove a very simple, but important
theorem for the quantum mechanics of higher derivative theories:
{\it Total derivative terms in a Lagrangian
never affect the quantum mechanics}.
The theorem is proven within the most general,
{\it i.e.} singular and non-autonomous (explicitly time-dependent), situation.
Needless to say, the classical version of the theorem is well known
and is trivial.
And the classical result often plays an important role in various theories.
Surprisingly enough, the quantum version, despite its importance,
has not been proven up to now.
It is in fact non-trivial and requires a proof.

Our plan is as follows:

In Sec.\ref{sec:general-th}, the canonical theory, now known as
the Ostrogradski formalism \cite{Gitman}, is reviewed and
the canonical quantization {\it \`{a} la} Dirac is performed.
Sec.\ref{sec:reduction-th} is devoted to the proof of the theorem.

\section{General theory of higher derivative systems}
\label{sec:general-th}

\subsection{The Ostrogradski formalism}
\label{subsec:Ostrogradski}

Let us consider a Lagrangian which depends on the coordinates $q^{i}$
and their time derivatives up to the $N$-th order,
   \begin{equation}
   L(\bq^{(0)},\bq^{(1)},\ldots,\bq^{(N)},t),               \label{Lagrangian}
   \end{equation}
where $\bq^{(I)}$'s are abbreviations for
$ q^{(I)i} := d^{I}q^{i} / dt^{I} $,     $ I=0,1,\ldots,N $.
The Euler-Lagrange equations for (\ref{Lagrangian}) are given by
   \begin{equation}
   \sum_{I=0}^{N} \left( -{d \over dt} \right)^I
   {\rd L \over \rd q^{(I)i}}=0.                        \label{Euler-Lagrange}
   \end{equation}
We introduce canonical variables,
   \begin{mathletters}                                       \label{variables}
   \begin{eqnarray}
   q^{Ii} &:=& q^{(I)i}, \label{q-variables} \\
   p_{Ii} &:=& \sum_{K=I+1}^N \left( -{d \over dt} \right)^{K-I-1}
   {\rd L \over \rd q^{(K)i}}, \label{p-varialbles}
   \end{eqnarray}
   \end{mathletters}
to parametrize the $2N$ dimensional (for each $i$) phase space of
(\ref{Euler-Lagrange}).
Throughout the present paper, we take the convension that $I$, $J$, and $K$
run from $0$ to $N-1$, unless otherwise stated.
There are relations as follows,
   \begin{mathletters}                                        \label{relation}
   \begin{eqnarray}
   & &{\rd L \over \rd q^{(0)i}} = {dp_{0i} \over dt}, \label{another-EL} \\
   & &p_{Ai} = {\rd L \over \rd q^{(A+1)i}} - {d \over dt}p_{A+1,i},
   \label{recursive}
   \end{eqnarray}
   \end{mathletters}
with $A=0,1,\ldots,N-2$.
Eqs.(\ref{another-EL}) are the Euler-Lagrange equations
(\ref{Euler-Lagrange}).

The Lagrangian (\ref{Lagrangian}) with
$ \det(\rd^2 L / \rd q^{(N)i} \rd q^{(N)j}) = 0 $
is singular, which means there exist primary constraints,
   \begin{equation}
   \phi_m(\bq,\bp,t)=0,                                    \label{constraints}
   \end{equation}
where $\bq$ and $\bp$ are abbreviations
for $q^{Ii}$'s and $p_{Ii}$'s.
The number of primary constraints is {\it larger than} or equal to
the nullity of the Hessian matrix
$ \| \rd^2 L / \rd q^{(N)i} \rd q^{(N)j} \| $;
the larger case may occur when $N \ge 2$.

The canonical Hamiltonian is defined by
   \begin{equation}
   H(\bq,\bp,t) :=
   \sum_{A=0}^{N-2} p_{Ai}q^{A+1,i}
   + p_{N-1,i} \dot{q}^{N-1,i}
   - L( \bq , \dot{\bq}^{N-1} , t ),                     \label{c-Hamiltonian}
   \end{equation}
which is conserved if the system is autonomous:
   \begin{equation}
   {dH \over dt} + {\rd L \over \rd t} = 0.               \label{conservation}
   \end{equation}
Note that the canonical Hamiltonian (\ref{c-Hamiltonian})
have some ambiguity as a function of $\bq$, $\bp$, and~$t$
because of the relations $q^{I+1,i}=\dot{q}^{Ii}$.
We fix the ambiguity by
distinguishing $q^{I+1,i}$'s and $\dot{q}^{Ii}$'s as appeared
in Eq.(\ref{c-Hamiltonian}).
With this distinction,
Hamilton's canonical equations of motion
become equivalent to the original
Euler-Lagrange equations (\ref{Euler-Lagrange}).
If we chose another form, we would obtain a Hamiltonian formalism
not equivalent to the original Lagrangian formalism.

Once we reach here, higher derivative theories do not differ
much from usual theories with first order derivatives.
The well-known Dirac procedure \cite{Gitman,Henneaux}
for singular Lagrangians is applied to higher derivative systems
without any modifications:

All the constraints (the primary and the secondary ones)
are classified into first-class, $\gamma_{a} \approx 0$,
and second class, $\chi_{\alpha} \approx 0$.
The Poisson bracket is defined by
   \begin{equation}
   \{ F,G \} := {\rd F \over \rd q^{Ii}}{\rd G \over \rd p_{Ii}}
   - {\rd F \over \rd p_{Ii}}{\rd G \over \rd q^{Ii}},      \label{Poisson-br}
   \end{equation}
with which the equations of motion for an arbitrary quantity
$F(\bq,\bp,t)$ is
   \begin{equation}
   \dot{F} \approx \{ F,H_{T} \} + {\rd F \over \rd t}.   \label{eq-of-motion}
   \end{equation}
The total Hamiltonian $H_{T}$ is defined as
   \begin{equation}
   H_{T} := H + u^{\alpha_{1}} \chi_{\alpha_{1}}
          + \lambda^{a_{1}} \gamma_{a_{1}},              \label{t-Hamiltonian}
   \end{equation}
where $H$ is the canonical Hamiltonian (\ref{c-Hamiltonian}),
$u^{\alpha_{1}}$'s are functions of $\bq$, $\bp$, and~$t$ determined
by consistency, and $\lambda^{a_{1}}$'s are
the arbitrary Lagrange multiplers.
The indeces
$\alpha_{1}$ and $a_{1}$ run only on the primary constraints.
The second-class constraints $\chi_{\alpha} \approx 0$
become strong equations $\chi_{\alpha} = 0$ in terms of
the Dirac bracket,
   \begin{equation}
   \{ F,G \}^{*} := \{ F,G \} - \{ F,\chi_{\alpha} \}
   C^{-1 \, \alpha \beta} \{ \chi_{\beta},G \},               \label{Dirac-br}
   \end{equation}
with $C_{\alpha \beta} := \{ \chi_{\alpha}, \chi_{\beta} \}$.

\subsection{Quantum mechanics}
\label{subsec:quantum}

The quantization is formally performed by replacing the Dirac bracket
$\{\; , \; \}^{*}$ by the quantum commutator
$(i \hbar )^{-1}[\; ,\; ]$.
Unfortunately, it is incredibly difficult to find out
the operator representation of the Dirac bracket.
In order to proceed further, it is desirable to circumvent
the Dirac bracket.
Here let us remember a general result \cite{Henneaux} that second-class
constraints can always be turned into first-class constraints,
if necessary, by adding extravariables.
Without losing generality, therefore, we assume
there is no second-class constraints.
Then we have only to consider the Poisson bracket.
Quantization is performed by replacing the Poisson bracket by
the commutator, and the first-class constraints by
the subsidiary conditions on the wave function.

Let us take the Schr\"{o}dinger picture with coordinate representation.
The commutator algebra is represented by
$ \hat{q}^{Ii} = q^{Ii} $, and
$ \hat{p}_{Ii} = -i\hbar {\rd / \rd q^{Ii} } $.
One obtains the Schr\"{o}dinger equation
   \begin{equation}
   i\hbar {\rd \psi (\bq ,t) \over \rd t}
   = \hat{H}
   \left(
     \bq , -i\hbar{\rd \over \rd \bq} , t
   \right)
   \psi (\bq,t),                                       \label{Schroedinger-eq}
   \end{equation}
and the subsidiary conditions
  \begin{equation}
  \hat{\gamma}_{a}
  \left(
    \bq , -i\hbar { \rd \over \rd \bq } ,t
  \right)
  \psi (\bq,t) = 0,                                    \label{subsidiary-cond}
  \end{equation}
in which $\bq$ and ${\rd / \rd \bq}$ are abbreviations
for $q^{Ii}$'s and ${\rd / \rd q^{Ii}}$'s.

Finally I finish this section by emphasizing peculiarity
of the quantum mechanics of higher derivative systems:
Since the coordinates and velocities are both treated
as coordinates in the Ostrogradski formalism,
the uncertainty principle does not work between them!

\section{The reduction theorem}
\label{sec:reduction-th}

\subsection{Problem setting}

Consider an $(N-1)$-th order Lagrangian $L^{\sharp}$ that may be
singular.  We define an $N$-th order Lagrangian $L$ as follows:
   \begin{equation}
   L(\bq^{(0)},\bq^{(1)},\ldots,\bq^{(N)},t) :=
   L^{\sharp}(\bq^{(0)},\bq^{(1)},\ldots,\bq^{(N-1)},t) +
   {d \over dt}W(\bq^{(0)},\bq^{(1)},\ldots,\bq^{(N-1)},t),       \label{Lagr}
   \end{equation}
where $W$ is an arbitrary function of $q^{(I)i}$'s and $t$.
Needless to say, $L$ and $L^{\sharp}$ are classically equivalent
because the total derivative term is turned, in the action integral,
into the surface term, which does not affect the classical
equations of motion.
Nevertheless their quantum equivalence is non-trivial.
Note that $L$ and $L^{\sharp}$ belong to defferent order
Lagrangians.
As is stated in Sec.\ref{sec:general-th}, different order Lagrangians
lead to different conjugate pairs, which means the different
uncertainty principle.
The canonical variables (\ref{variables}) for the Lagrangian
(\ref{Lagr}) is given as
   \begin{mathletters}                                            \label{vari}
   \begin{eqnarray}
   & &q^{Ii} = q^{(I)i}, \label{q's} \\
   & &p_{N-1,i} = {\rd W \over \rd q^{(N-1)i}}, \label{p(N-1)} \\
   & &p_{Ai} = p^{\sharp}_{Ai} + {\rd W \over \rd q^{(A)i}}. \label{p(A)}
   \end{eqnarray}
   \end{mathletters}
We take the convention that $I$, $J$, $K$ run from $0$ to $N-1$,
and $A$, $B$, $C$ run from $0$ to $N-2$, unless otherwise stated.
Here $p^{\sharp}_{Ai}$'s are the canonical momentum
in the $L^{\sharp}$-theory,
   \begin{mathletters}                                         \label{p-sharp}
   \begin{eqnarray}
   & &p^{\sharp}_{N-2,i} := {\rd L^{\sharp} \over \rd q^{(N-1)i}},
      \label{p-sharp(N-1)} \\
   & &p^{\sharp}_{Ai} := {\rd L^{\sharp} \over \rd q^{(A+1)i}}
      - {d \over dt}p^{\sharp}_{A+1,i},
      \label{p-sharp(A)}
   \end{eqnarray}
   \end{mathletters}
with $A=0,1,\ldots,N-3$.
The canonical Hamiltonian (\ref{c-Hamiltonian}) becomes
   \begin{eqnarray}
   H(\bq,\bp,t)
   &=& p_{Ai}q^{A+1,i} + p_{N-1,i}\dot{q}^{N-1,i}
       - L( \bq, \dot{\bq}^{N-1}, t ) \nonumber \\
   &=& H^{\sharp}(\bq,\bp^{\sharp},t)
       - {\rd W(\bq,t) \over \rd t},                            \label{c-Hami}
   \end{eqnarray}
with
   \begin{equation}
   H^{\sharp}(\bq,\bp^{\sharp},t)
   := p^{\sharp}_{Ai}q^{A+1,i} - L^{\sharp}(\bq,t).         \label{sharp-Hami}
   \end{equation}
We should distinguish
$H^{\sharp}(\bq,\bp^{\sharp},t)$ defined here
from the canonical Hamiltonian in the $L^{\sharp}$-theory,
   \begin{equation}
   H^{\sharp}(\bq_{\sharp},\bp^{\sharp},t)
   := \sum_{A=0}^{N-3} p^{\sharp}_{Ai}q^{A+1,i}
   + p^{\sharp}_{N-2,i}\dot{q}^{N-2,i}
   - L^{\sharp}(\bq_{\sharp},\dot{\bq}^{N-2},t),           \label{sharp-Hami2}
   \end{equation}
where $\bq_{\sharp}$ is an abbreviation for $q^{Ai}$'s,
while $\bq$ without $\sharp$ is the abbreviation for $q^{Ii}$'s.

These two $H^{\sharp}$'s, (\ref{sharp-Hami}) and (\ref{sharp-Hami2}),
are the same in their value but different
as functions of canonical variables.  This implies the following:
While Eq.(\ref{c-Hami}) with Eq.(\ref{sharp-Hami})
defines a Hamiltonian system equivalent to
the original Lagrangian system (\ref{Lagr}),
the use of Eq.(\ref{sharp-Hami2})
instead of Eq.(\ref{sharp-Hami})
results in another Hamiltonian system {\it not} equivalent to
the original Lagrangian system.
In other words, Eq.(\ref{sharp-Hami2}) have forgotten the relations
   \begin{equation}
   \dot{q}^{N-2,i} = q^{N-1,i}.                              \label{forgotten}
   \end{equation}
Eq.(\ref{sharp-Hami}), on the other hand, remember them
as canonical equations,
$ \dot{q}^{N-2,i} = \rd H(\bq,\bp,t) / \rd p_{N-2,i} $.
Thus Eq.(\ref{sharp-Hami2}) defines a larger theory,
in which the original one is contained as a special case.
If we impose, by hand, the relation (\ref{forgotten})
on the larger theory,
then it reduces to the original one.

Since the larger theory is quite useful for our purposes,
we use, in the following,
(\ref{sharp-Hami2}) instead of (\ref{sharp-Hami}).
As will be stated in Sec.\ref{subsec:gauge-transf},
the larger theory is a kind of gauge theory and
we can impose Eq.(\ref{forgotten}) as a gauge fixing condition
to it.

\subsection{Constraint analysis}

Equations (\ref{p(N-1)}) are primary constraints:
   \begin{equation}
   \gamma_{i}(\bq,\bp,t) := p_{N-1,i} -
   {\rd W \over \rd q^{N-1,i}} \approx 0.                            \label{g}
   \end{equation}
When $L^{\sharp}$ is singular, {\it i.e.}
$\det (\rd^{2}L^{\sharp} / \rd q^{(N-1)i} q^{(N-1)j}) = 0$,
there are other primary constraints in addition to $\gamma_{i}$'s:
   \begin{equation}
   \gamma^{\sharp}_{a}(\bq_{\sharp},\bp^{\sharp},t)
   \approx 0,                                                   \label{gsharp}
   \end{equation}
which stem from (\ref{p(A)}) together with (\ref{p-sharp}).
It is not a hard task to show
   \begin{mathletters}                                      \label{g-g-gsharp}
   \begin{eqnarray}
   \{\gamma_{i},\gamma_{j}\}
   &=& - \left\{
         p_{N-1,i} , { \rd W \over \rd q^{N-1,j} }
         \right\}
       - \left\{
         { \rd W \over \rd q^{N-1,i} } , p_{N-1,j}
         \right\} \nonumber \\
   &=&   { \rd^{2}W \over \rd q^{N-1,i} \rd q^{N-1,j} }
       - { \rd^{2}W \over \rd q^{N-1,j} \rd q^{N-1,i} } \nonumber \\
   &=& 0, \label{g-g} \\
   \{\gamma_{i},\gamma^{\sharp}_{a}\}
   &=& \{ p_{N-1,i} , q^{N-1,k} \}
       { \rd\gamma^{\sharp}_{a} \over \rd p^{\sharp}_{Aj} }
       { \rd p^{\sharp}_{Aj} \over \rd q^{N-1,k} }
       -
       \left\{ { \rd W \over \rd q^{N-1,i} } , p_{Bk} \right\}
       { \rd\gamma^{\sharp}_{a} \over \rd p^{\sharp}_{Aj} }
       { \rd p^{\sharp}_{Aj} \over \rd p_{Bk} } \nonumber \\
   &=& { \rd\gamma^{\sharp}_{a} \over \rd p^{\sharp}_{Aj} }
       { \rd^{2}W \over \rd q^{N-1,i} \rd q^{Aj} }
       -
       { \rd^{2}W \over \rd q^{A,j} \rd q^{N-1,i} }
       { \rd\gamma^{\sharp}_{a} \over \rd p^{\sharp}_{Aj} }
       \nonumber \\
   &=& 0. \label{g-gshap}
   \end{eqnarray}
   \end{mathletters}
The total Hamiltonian is given by
   \begin{equation}
   H_{T} = H + \lambda^{a}\gamma^{\sharp}_{a}
   + \lambda^{i}\gamma_{i},                                     \label{t-Hami}
   \end{equation}
where $\lambda^{a}$'s and $\lambda^{i}$'s are
the Lagrange multipliers.
Straightforward calculation shows that
$\gamma_{i}$'s do not produce secondary constraints:
   \begin{eqnarray}
   \dot{\gamma}_{i}
   & \approx &
         \{ \gamma_{i} , H_{T} \} + { \rd\gamma_{i} \over \rd t}
         \nonumber \\
   & \approx &
         \left\{ \gamma_{i} , H^{\sharp} - {\rd W \over \rd t} \right\}
         + { \rd\gamma_{i} \over \rd t }
         \nonumber \\
   & = & { \rd H^{\sharp} \over \rd p^{\sharp}_{Aj} }
         { \rd^{2}W \over \rd q^{N-1,i} \rd q^{Aj} }
       + { \rd^{2}W \over \rd q^{N-1,i} \rd t }
       - { \rd^{2}W \over \rd q^{Aj} \rd q^{N-1,i} }
         { \rd H^{\sharp} \over \rd p^{\sharp}_{Aj} }
       - { \rd^{2}W \over \rd t \, \rd q^{N-1,i} }
         \nonumber \\
   & = & 0.                                                      \label{g-dot}
   \end{eqnarray}
Therefore one concludes that $\gamma_{i}$'s are first-class.
This conclusion were not obtained if we would adopt
Eq.(\ref{sharp-Hami}) in Eq.(\ref{c-Hami}).
As for $\gamma^{\sharp}_{a}$'s, one obtains
   \begin{eqnarray}
   \dot{\gamma}^{\sharp}_{a}
   & \approx &
         \{ \gamma^{\sharp}_{a} , H_{T}  \}
       + { \rd\gamma^{\sharp}_{a} \over \rd t }
         \nonumber \\
   & \approx &
         \{ \gamma^{\sharp}_{a} , H^{\sharp}  \}
       + { \rd\gamma^{\sharp}_{a} \over \rd p^{\sharp}_{Ai} }
         { \rd^{2}W \over \rd q^{Ai} \rd t }
       + \lambda^{b}
         \{ \gamma^{\sharp}_{a} , \gamma^{\sharp}_{b} \}
       - { \rd\gamma^{\sharp}_{a} \over \rd p^{\sharp}_{Ai} }
         { \rd^{2}W \over \rd t \, \rd q^{Ai} }
       + \left(
         { \rd\gamma^{\sharp}_{a} \over \rd t }
         \right)_{\sharp}
         \nonumber \\
   & \approx &
         \{ \gamma^{\sharp}_{a} , H^{\sharp}_{T}  \}
       + \left(
         { \rd\gamma^{\sharp}_{a} \over \rd t }
         \right)_{\sharp},                             \label{gsharp-dot-temp}
   \end{eqnarray}
where
   \begin{equation}
   H^{\sharp}_{T}
   := H^{\sharp} + \lambda^{a}\gamma^{\sharp}_{a}         \label{sharp-t-Hami}
   \end{equation}
is the total Hamiltonian in the $L^{\sharp}$-theory.
The symbol
$(\rd / \rd t)_{\sharp}$
represents partial derivative by $t$ with
$\bq_{\sharp}$ and $\bp^{\sharp}$ fixed.
One can prove
   \begin{mathletters}                                   \label{fund-sharp-br}
   \begin{eqnarray}
   & &\{ q^{Ai} , p^{\sharp}_{Bj} \}
   = \delta^{A}_{B}\delta^{i}_{j}, \\
   & &\{ q^{Ai} , q^{Bj} \}
   = \{ p^{\sharp}_{Ai} , p^{\sharp}_{Bj} \}
   = 0,
   \end{eqnarray}
   \end{mathletters}
which mean
   \begin{equation}
   \{ F^{\sharp} , G^{\sharp} \}
   = \{ F^{\sharp} , G^{\sharp} \}_{\sharp},               \label{br-relation}
   \end{equation}
where $F^{\sharp}$ and $G^{\sharp}$ are arbitrary functions
of $\bq_{\sharp}$, $\bp^{\sharp}$, and~$t$. The symbol
$\{ \; , \; \}_{\sharp}$ is the Poisson bracket
in the $L^{\sharp}$-theory.
Using (\ref{br-relation}),
one can rewrite (\ref{gsharp-dot-temp}) as
   \begin{equation}
   \dot{\gamma}^{\sharp}_{a} \approx
   \{ \gamma^{\sharp}_{a} , H^{\sharp}_{T}  \}_{\sharp}
       + \left(
         { \rd\gamma^{\sharp}_{a} \over \rd t }
         \right)_{\sharp}.                                  \label{gsharp-dot}
   \end{equation}
Note that
the only property we assumed in deriving the above equations is
that $\gamma^{\sharp}_a$'s are functions of
$\bq_{\sharp}$, $\bp^{\sharp}$, and~$t$.
Thus Eqs.(\ref{gsharp-dot}) remain valid even if
secondary constraints are substituted for
$\gamma^{\sharp}_{a}$'s.
Thus we have proven that
{\it all the secondary constraints emerge from
$\gamma^{\sharp}_{a}$'s
are just the same as the ones derived in the
$L^{\sharp}$-theory.}

As is stated in Sec.\ref{subsec:quantum},
in the following, we do assume that
$\gamma^{\sharp}_{a}$'s
and the secondary constraints derived from them are
all first-class.
Let us write them again as
$\gamma^{\sharp}_{a} \approx 0$.
Then all the constraints in our Lagrangian (\ref{Lagr})
are exhausted by (\ref{g})
and (\ref{gsharp}).
The final form of the total Hamiltonian is (\ref{t-Hami}),
in which the summation on $a$ should be taken
only on the primary constraints.
(The summation on all the first-class constraints defines
the extended Hamiltonian formalism \cite{Henneaux};
our discussion in what follows
remains valid even if we take the extended formalism.)

\subsection{Gauge transformations}
\label{subsec:gauge-transf}

In this subsection we investigate the gauge transformation
derived from $\gamma_{i}$'s.
For an arbitrary quantity
$F(\bq,\bp,t)$, the gauge transformation
is defined as
   \begin{equation}
   \delta F := \varepsilon^{i}
   \{ F , \gamma_{i} \},                                  \label{gauge-transf}
   \end{equation}
where $\varepsilon^{i}$'s are arbitrary functions of~$t$,
but are independent of the canonical variables.
As is well known, physical quantities must be gauge invariant.

One can show that quantities related to the $L^{\sharp}$-theory,
for example $q^{Ai}$, $p^{\sharp}_{Ai}$, and $H^{\sharp}$,
are all gauge invariant.
Whereas quantities proper to the $L$-theory are, in general,
non-invariant.
For example, one obtains
   \begin{mathletters}                                    \label{gauge-noninv}
   \begin{eqnarray}
   &\delta& p_{Ii}
          = \varepsilon^{j}
          { \rd^{2}W \over \rd q^{Ii} \rd q^{N-1,j} }, \\
   &\delta& q^{N-1,i}
          = \varepsilon^{i}, \\
   &\delta& W
          = \varepsilon^{i}
          { \rd W \over \rd q^{N-1,i} }, \\
   &\delta& H_{T}
          = \delta H
          = - \varepsilon^{i}
          { \rd^{2}W \over \rd t \, \rd q^{N-1,i} },
   \end{eqnarray}
   \end{mathletters}
which shows that $p_{Ii}$'s, $q^{N-1,i}$'s, $W$,
$H_{T}$, and $H$ are all non-invariant and unphysical.

Needless to say, true physical quantities must be
gauge invariant under the gauge transformations
derived from $\gamma^{\sharp}_{a}$'s as well.
Further investigation of them requires the specification of
$L^{\sharp}$'s concrete form.
So we do not pursue it any more.

Notice:
Remember that our theory is
equivalent to the original Lagrangian system (\ref{Lagr})
when we impose Eqs.(\ref{forgotten}) in addition.
Since Eqs.(\ref{forgotten}) are not gauge invariant,
they work as a gauge fixing condition.
The gauge transformations derived from $\gamma_{i}$'s are
the symmetry of the larger theoy but
not the symmetry of the original Lagrangian (\ref{Lagr}).
Nevertheless they are important; gauge invariance of our theory makes it
clear that the conditions (\ref{forgotten}) are in fact unessential and
do not affect the physics.

\subsection{Proof of the theorem}

We now turn our attention to the quantum mechanics.
The Schr\"{o}dinger equation is given by (\ref{Schroedinger-eq})
with the Hamiltonian operator,
   \begin{equation}
   \hat{H}
      \left(
      \bq, -i\hbar { \rd \over \rd \bq }, t
      \right)
   = \hat{H}^{\sharp}
      \left(
      \bq_{\sharp} ,
      -i\hbar { \rd \over \rd \bq_{\sharp} }
      - { \rd W \over \rd \bq_{\sharp} } , t
      \right)
   - { \rd W \over \rd t },                                    \label{Hami-op}
   \end{equation}
derived from (\ref{c-Hami}) with (\ref{sharp-Hami2}) and (\ref{p(A)}).
The subsidiary conditions (\ref{subsidiary-cond}) are given by
   \begin{eqnarray}
   \hat{\gamma}_{i}\psi(\bq,t)
   = \left(
           -i\hbar{ \rd \over \rd q^{N-1,i} }
           - { \rd W \over \rd q^{N-1,i} }
     \right)
   \psi(\bq,t) = 0,                                             \label{g-subs}
   \\
   \hat{\gamma}^{\sharp}_{a}
   \left(
         \bq_{\sharp} ,
         -i\hbar{ \rd \over \rd \bq_{\sharp} }
         - { \rd W \over \rd \bq_{\sharp} } , t
   \right)
   \psi(\bq,t) = 0.                                        \label{gsharp-subs}
   \end{eqnarray}
Eqs.(\ref{g-subs}) are solved as follows:
   \begin{equation}
   \psi(\bq,t) =
   \psi^{\sharp}(\bq_{\sharp},t)
   \exp{ iW(\bq,t) \over \hbar },                           \label{phys-state}
   \end{equation}
where $\psi^{\sharp}$ is the arbitrary function of
$q^{Ai}$'s and~$t$.
Eq.(\ref{phys-state}) gives the general form of the physical state.
Note that $\psi^{\sharp}(\bq_{\sharp},t)$ is gauge invariant,
while $\psi(\bq,t)$ is not.

It is not hard to verify the identities
   \begin{equation}
   \left(
   -i\hbar { \rd \over \rd q^{Ai} }
   - { \rd W \over \rd q^{Ai} }
   \right)^{n} \psi
   =
   \left[ \left(
   -i\hbar { \rd \over \rd q^{Ai} }
   \right)^{n} \psi^{\sharp}
   \right]
   \exp {iW \over \hbar},                                     \label{identity}
   \end{equation}
on the physical state (\ref{phys-state}).
Here $n$ is a non-negative integer.
These identities imply the following identity:
   \begin{equation}
   \hat{F}
   \left(
   \bq_{\sharp} ,
   -i\hbar { \rd \over \rd \bq_{\sharp} }
   - { \rd W \over \rd q^{Ai} } , t
   \right) \psi
   =
   \left[
   \hat{F}
   \left(
   \bq_{\sharp} ,
   -i\hbar { \rd \over \rd \bq_{\sharp} } , t
   \right) \psi^{\sharp}
   \right]
   \exp { iW \over \hbar }.                                       \label{iden}
   \end{equation}
Here
$F(\bq_{\sharp},\bp^{\sharp},t)$
is an arbitrary quantitiy
which is a polynomial with respect to
$p^{\sharp}_{Ai}$'s.
Assuming that
$H^{\sharp}(\bq_{\sharp},\bp^{\sharp},t)$ and
$\gamma^{\sharp}_{a}(\bq_{\sharp},\bp^{\sharp},t)$'s
are polynomials with respect to
$p^{\sharp}_{Ai}$'s,
we apply (\ref{iden}) for them.

Inserting (\ref{phys-state}) into the Schr\"{o}dinger
equation (\ref{Schroedinger-eq}) with (\ref{Hami-op}),
and using (\ref{iden}), we finally obtain
   \begin{equation}
   i\hbar
   { \rd \psi^{\sharp}(\bq_{\sharp},t) \over \rd t }
   =
   \hat{H}^{\sharp}
   \left(
   \bq_{\sharp} ,
   -i\hbar { \rd \over \rd \bq_{\sharp} } , t
   \right)
   \psi^{\sharp}(\bq_{\sharp},t).                               \label{r-Schr}
   \end{equation}
This is nothing but the Schr\"{o}dinger equation in the
$L^{\sharp}$-theory.
As for the condition (\ref{gsharp-subs}), it simply becomes
   \begin{equation}
   \hat{\gamma}^{\sharp}_{a}
   \left( \bq_{\sharp} ,
   -i\hbar { \rd \over \rd \bq_{\sharp} } , t
   \right)
   \psi^{\sharp}(\bq_{\sharp},t) = 0.                    \label{r-gsharp-subs}
   \end{equation}
Eq.(\ref{r-Schr}) together with Eqs.(\ref{r-gsharp-subs}) constitutes
the full quantum mechanics for the $L^{\sharp}$-theory.
Here $\psi^{\sharp}(\bq_{\sharp},t)$ is identified as the wave function
of the $L^{\sharp}$-theory.

We have proven the following proposition:

\begin{proposition}                                        \label{proposition}
   $n$-th order and $(n-1)$-th order Lagrangians,
   which may be singular and non-autonomous,
   lead to the same quantum mechanics
   if their difference is a total time derivative.
\end{proposition}

Proposition \ref{proposition} implies the following theorem:

\begin{theorem}                                                \label{theorem}
   $n$-th order and $m$-th order Lagrangians,
   which may be singular and non-autonomous,
   lead to the same quantum mechanics
   if their difference is a total time derivative.

Proof:\rm \,
Let us assume $n \ge m$, and put
   \begin{equation}
   L_{n} = L^{\sharp}_{m}
         + { d \over dt } W_{n-1}.                                  \label{Ln}
   \end{equation}
Here, the subscript of $L$, $L^{\sharp}$, and $W$ denote
the order of highest derivatives contained.
We use the same notation for ${\cal L}$ and ${\cal W}$,
which appear in what follows, as well.

\begin{enumerate}
\item $n = m$ case

Let us introduce an arbitrary function ${\cal W}_{n}$,
and define a new Lagrangian ${\cal L}_{n+1}$
as follows:
   \begin{equation}
   {\cal L}_{n+1} :=
   L_{n} + { d \over dt }{\cal W}_{n}.
   \end{equation}
Then (\ref{Ln}) is rewritten as
   \begin{equation}
   {\cal L}_{n+1} = L^{\sharp}_{n} +
   { d \over dt }(W_{n-1} + {\cal W}_{n}).
   \end{equation}
These equations together with Proposition \ref{proposition} say
   \begin{equation}
   L_{n} \sim {\cal L}_{n+1} \sim L^{\sharp}_{n},
   \end{equation}
where $\sim$ means quantum mechanically equivalent.

\item $n = m+1$ case

This case is just the same as Proposition \ref{proposition}.

\item $n \ge m+2$ case

Let us introduce $n-m-1$ arbitrary functions ${\cal W}_{i}$,
$ i = m , m+1 , \ldots , n-2 $,
and define the same number of new Lagrangians ${\cal L}_{i}$,
$ i = m+1 , m+2 , \ldots , n-1 $,
as follows:
   \begin{eqnarray}
   {\cal L}_{m+1}
   &:=& L^{\sharp}_{m} + { d \over dt }{\cal W}_{m}, \nonumber \\
   {\cal L}_{m+2}
   &:=& {\cal L}_{m+1} + { d \over dt }({\cal W}_{m+1} - {\cal W}_{m}), \\
   & \ldots & \ldots \nonumber \\
   {\cal L}_{n-1}
   &:=& {\cal L}_{n-2} + { d \over dt }({\cal W}_{n-2} - {\cal W}_{n-3}).
   \nonumber
   \end{eqnarray}
Consistency between these equations and (\ref{Ln}) requires
   \begin{equation}
   L_{n} = {\cal L}_{n-1}
         + { d \over dt }(W_{n-1} - {\cal W}_{n-2}).
   \end{equation}
Therefore we have proven
   \begin{equation}
   L^{\sharp}_{m} \sim {\cal L}_{m+1}
                  \sim {\cal L}_{m+2}
                  \sim \cdots
                  \sim {\cal L}_{n-2}
                  \sim {\cal L}_{n-1}
                  \sim L_{n}.
   \end{equation}
This completes the proof.
\hfill $\Box$
\end{enumerate}
\end{theorem}

\begin{corollary}[Grosse-Knetter]                            \label{corollary}
An $m$-th order Lagrangian, and the same Lagrangian formally
treated as if it were an $n(\ge m)$-th order Lagrangian
lead to the same quantum mechanics.

Proof:\rm \,
This is the special case of Theorem \ref{theorem}
with the vanishing total derivative term.
\hfill $\Box$
\end{corollary}

Corollary \ref{corollary} has been proven, using the path integral,
by Grosse-Knetter \cite{Grosse-Knetter}, for the special case of
autonomous Lagrangians.
Our Corollary \ref{corollary} is the generalization of his result
to non-autonomous Lagrangians.

\acknowledgments
The author thanks K. Kamimura (Toho University)
and J. Gomis (Universitat de Barcelona) for discussions.


\end{document}